\documentclass[aps,prf]{revtex4-1}
\usepackage{amsmath,amsfonts,bm, color,amssymb}
\usepackage{graphicx,epsfig,overpic,hyperref,ulem}
\input epsf
\UseRawInputEncoding

\newcommand{\eps}{{\varepsilon}}
\newcommand{\sigm}{\sigma}
\newcommand{\out}{{\mathrm{s}}}
\newcommand{\ins}{{\mathrm{d}}}
\newcommand{\ehd}{{\mathrm{ehd}}}
\newcommand{\dep}{{\mathrm{dep}}}
\newcommand{\el}{{\mathrm{el}}}

\newcommand{\visrat}{{\lambda}}
\newcommand{\eq}{{\mathrm{eq}}}

\newcommand{\bs}{\boldsymbol}
\newcommand{\Rr}{\mbox{\it R}}
\newcommand{\Sr}{\mbox{\it S}}
\newcommand{\bE}{{\bf E}}
\newcommand{\bF}{{\bf F}}
\newcommand{\bT}{{\bf T}}
\newcommand{\bU}{{\bf U}}
\newcommand{\bI}{{\bf I}}
\newcommand{\bP}{{\bf P}}
\newcommand{\bR}{{\bf \hat d}}
\newcommand{\bu}{{\bs u}}
\newcommand{\bff}{{\bf f}}
\newcommand{\bt}{{\bf{\hat{t}}}}

\newcommand{\xhat}{{\bf \hat x}}

\newcommand{\zhat}{{\bf \hat z}}
\newcommand{\rhat}{{\bf \hat r}}
\newcommand{\that}{{\bm \hat \theta}}
\newcommand{\bx}{{\bf x}}
\newcommand{\by}{{\bf y}}

\newcommand{\bn}{{\bf n}}
\newcommand{\bnabla}{{\bm \nabla}}

\newcommand{\Ca}{\mbox{\it Ca}}
\newcommand{\Pe}{\mbox{\it Pe}}
\newcommand{\Ma}{\mbox{\it Ma\,}}
\newcommand{\E}{\mbox{\it E}}
\newcommand{\half}{\frac{1}{2}}

\newcommand{\refeq}[1]{Eq. (\ref{#1})}

\begin{document}

\title{Note on the pairwise interactions of surfactant-covered drops in a uniform electric field}
\author{Chiara Sorgentone$^1$  and Petia M. Vlahovska$^2$}
\affiliation{ $^1$ Department of Basic and Applied Sciences for Engineering, Sapienza Universit\`a di Roma, 00161 Rome, Italy
\\
$^2$ Engineering Sciences and Applied Mathematics, Northwestern University, Evanston, IL 60208, USA}

\begin{abstract}
We study the effect of surfactant on the  pairwise interactions of drops in an applied uniform DC electric field using a combination of numerical simulations based on a boundary integral formulation and an analytical theory assuming small drop deformations.  The surfactant is assumed to be insoluble in the bulk-phase fluids.
We show that the surfactant weakens the electrohydrodynamic flow and thus dielectrophoretic interactions play more prominent role in the  dynamics of surfactant-covered drops compared to clean drops.  If drop conductivity is the same as the suspending fluid, a nondiffusing  surfactant can arrest the drops' relative motion thereby effectively preventing coalescence.

\end{abstract}

\date{preprint}

\maketitle

\section{Introduction}
\label{sec:intro}
Electric fields are widely used to manipulate particles and fluids. 
For example, separation of emulsified water from crude oil  in the petroleum refining process is achieved by the application of electric fields, which facilitate drop coalescence \citep{Waterman,Eow:2002}. 
An important  question pertains to the influence of surface-active substances (surfactants, compounds that lower the surface tension between liquids), which are naturally present in the crude oil (asphaltenes, resins, acids), on the process droplet attraction and coalescence.

The effect of surfactants and electric fields on drop dynamics has been largely studied in two-dimensions, whilst in three-dimensions the literature is limited. The effect of surfactant (no electric field) has been studied using simulations based on the boundary integral method \citep{Li-Pozrikidis:1997,Pozrikidis:2004,Stone-Leal:1990,Yon-Pozrikidis:1998,Eggleton-Tsai-Stebe:2001,Bazhlekov:2006,Feigl:2007, Vlahovska:2005,Rother:2006}, the diffuse-interface-method \citep{Teigen:2011}, a front-tracking method \citep{Muradoglu-Tryggvason:2008} or a conserving volume-of-fluid method \citep{James-Lowengrub:2004}. 
The effect of electric fields on clean drops (no surfactant) has been studied theoretically, numerically and experimentally both for single and multiple drops \citep{Lac-Homsy, Karyappa-Thaokar:2014, Lanauze:2015, Ha:2000a, DavidS:2017b, Fernandez:2008a, Casas:2019, Baygents:1998, Lin:2012, Mhatre:2015, Salipante-Vlahovska:2010, sozou_1975, Zabarankin}, 
and we refer the interested reader to our recent work \citep{Chiara:2020} for a more extensive bibliography. In that paper we presented a detailed analysis of the three-dimensional  interaction of a drop pair in a uniform electric field; we showed that the pair dynamics are not simple attraction or repulsion; depending on the angle between the center-to-center line with the undisturbed electric field, the relative motion of the two particles can be quite complex. For example, they can attract in the direction of the field and move towards each other, pair up, and then separate in the transverse direction. 

The combined effect of surfactants and electric fields is a  virtually unexplored problem in terms of numerical experiments, especially when considering multiple drops. This is due to the numerous computational challenges associated with the complex moving geometries and the multi-physics nature of the problem. Teigen and Munkejord used a level-set method in an axisymmetric, cylindrical coordinate system to investigate the interaction between surfactant-covered drops with a uniform electric field for a single drop and, more recently, Poddar et al., studied theoretically the electrorheology of a dilute emulsion of surfactant-covered drops \citep{poddar_mandal_bandopadhyay_chakraborty_2019}. Other theoretical studies developed asymptotic analyses \citep{Ha:1995,Herve:2013,PhysRevE.99.063104} to investigate the deformation and the effects of surfactant transport on the deformation of a single viscous drop under a DC electric field.

In  this note we built upon our previous work  \citep{Chiara:2019, Chiara:2020} and explore the effect of an insoluble surfactant on a drop pair electrohydrodynamics. 

\section{Problem formulation}
\begin{figure}
\centerline{\includegraphics[width=.70\linewidth]{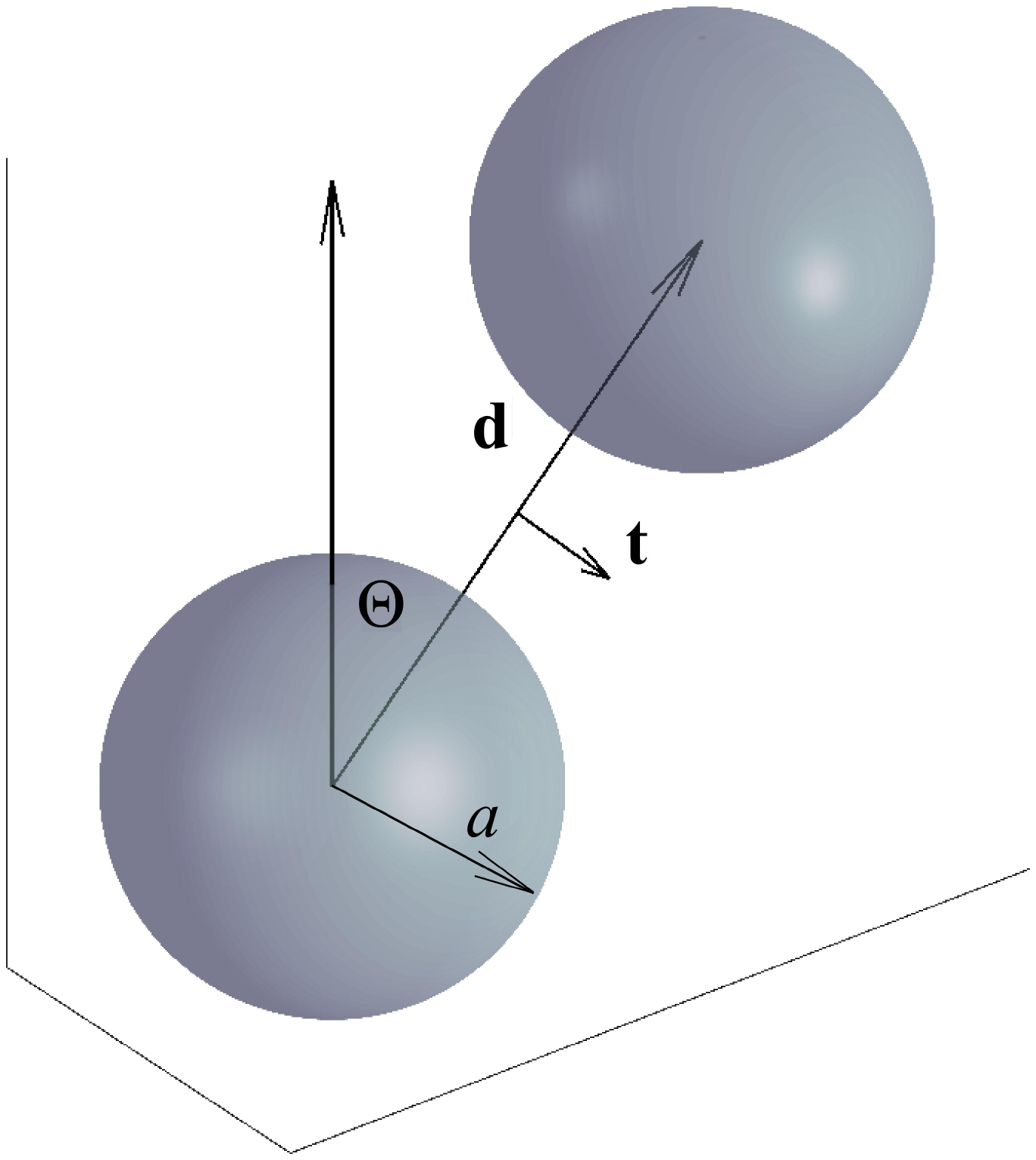}}
   \begin{picture}(0,0)(0,0)
\put(-130,100){\rotatebox{90}{{\Large$\rightarrow$}}$\bE^\infty$}
\put(0,20){$x$}
\put(-100,25){$y$}
\put(-100,150){$z$}
\put(-25,70){$1$}
\put(15,180){$2$}
\end{picture}
\caption{\footnotesize{ Two initially spherical identical drops with radius $a$, permittivity $\eps_\ins$ and conductivity $\sigm_\ins$ suspended in a fluid  permittivity $\eps_\out$ and conductivity $\sigm_\out$ and subjected to a uniform DC electric field $\bE^\infty=E_0\zhat$. {The angle between the line-of-centers vector and the field direction is $\Theta=\arccos(\zhat\cdot\bR)$.}}}
\label{fig1}
\end{figure}

Let us consider 
two identical neutrally-buoyant and charge-free drops  with radius $a$, viscosity $\eta_\ins$, conductivity  $\sigm_\ins$, and permittivity  $\eps_\ins$ suspended in a fluid with viscosity $\eta_\out$,  conductivity $\sigm_\out$, and  permittivity  $\eps_\out$. 
The mismatch in drop and suspending fluid  properties is characterized by the conductivity, permittivity, and viscosity ratios 
\begin{equation}
\Rr=\frac{\sigm_\ins}{\sigm_\out}\,,\quad \Sr=\frac{\eps_\ins}{\eps_\out}\,,\quad \lambda=\frac{\eta_\ins}{\eta_\out}\,.
\end{equation}
 A monolayer of insoluble
 surfactant is adsorbed on the drop interfaces.  At rest, the
 surfactant distribution is uniform and the equilibrium surfactant
 concentration is $\Gamma_\eq$; the corresponding interfacial tension
 is $\gamma_\eq$.  
The distance between the drops' centroids is $d$ and the angle between the drops' line-of-centers with the applied field direction is $\Theta$. The  unit separation
vector between the  drops is defined by the difference between the position vectors of the drops' centers of mass $\bR=(\bx_2^c-\bx^c_1)/d$. The unit vector normal to the drops line-of-centers and orthogonal to $\bR$ is $\bt$.
The problem geometry is sketched in Figure \ref{fig1}.

We adopt the leaky dielectric model~\citep{Melcher-Taylor:1969}, which assumes creeping flow and charge-free bulk fluids acting as Ohmic conductors.
The assumption of charge-free fluids decouples the electric and hydrodynamic fields in the bulk. Accordingly,
\begin{equation}
\label{stress_bulk}
\eta \nabla^2\bu-\nabla p=0\,,\quad \nabla\cdot \bE=0\,,
\end{equation}
where   $\bu$ and $p$ are the fluid velocity  and pressure, and $\bE$ is the electric field. Far away from the drops, $\bE^\out\rightarrow \bE^\infty=E_0\zhat$ and $\bu\rightarrow 0$.

 The coupling  of the electric field and the fluid flow 
occurs at the drop interfaces $\cal{D}$, where the charges brought by conduction accumulate.
The Gauss' law dictates that  while the electric field  in the electroneutral bulk  fluids is solenoidal, 
 at the drop interface the electric displacement field, $\eps \bE$, is discontinuous and its jump corresponds to the surface charge density
 \begin{equation}
\eps\left(E_n^\out-\Sr E_n^\ins\right)=q\,,  \quad \bx\in \cal{D}
 \end{equation}
 where $E_n=\bE\cdot\bn$, 
 and  $\bn$ is the outward pointing normal vector to the drop interface. 
The surface charge density adjusts to satisfy the current balance
 \begin{equation}
\label{currencond1}
\frac{\partial q}{\partial t}+\nabla_s\cdot\left(\bu q\right) =\sigm_\out \left(E_n^\out-\Rr E_n^\ins\right)\,,  \quad \bx\in \cal{D}\,.
  \end{equation}
{In this study, we  neglect charge relaxation and convection, thereby reducing the charge conservation equation  to continuity of the electrical current
across the interface as originally proposed by \cite{Taylor:1966}}
\begin{equation}
\label{currencond}
E_n^\out=\Rr E_n^\ins\,.
\end{equation}
This simplification implies $\eps^2_\out E_0^2/(\eta_\out\sigma_\out)\ll1$. This condition is satisfied for the typical fluids used in experiments such as castor oil (conductivity is $\sim 10^{-11}$ S/m,  viscosity is $\sim 1$ Pa.s) and low field strengths  $E_0\sim 10^4$ V/m. 
  
 The electric field acting on the induced surface charge gives rise to electric shear stress at the interface. The tangential stress balance yields
 \begin{equation}
\label{stress balanceT}
\left(\bI-\bn\bn\right)\cdot \left( \bT^\out- \bT^\ins\right)\cdot\bn+q\bE_t=-\bnabla_s\gamma \,, \quad \bx\in \cal{D}\,,
\end{equation}
where  $T_{ij}=-p\delta_{ij}+\eta (\partial_j u_i+\partial_i u_j)$ is the hydrodynamic stress
  and $\delta_{ij}$ is the Kronecker delta function. The electric tractions is calculated from the Maxwell stress tensor $T^\el_{ij}=\eps \left(E_iE_j-E_kE_k\delta_{ij}/2\right)$.  $\gamma$ is the interfacial tension, which depends on the local surfactant
concentration $\Gamma$. $\bE_t=\bE-E_n\bn$ is the  tangential component of the electric field, which is continuous across the interface,  and $\bI$ is the idemfactor. The normal stress balance is
\begin{equation}
\label{stress balance}
\bn \cdot\left( \bT^\out- \bT^\ins\right)+\half\left(\left(E_n^{\out}\right)^2-\Sr \left(E_n^{\ins}\right)^2-(1-\Sr)E_t^2\right)=\gamma\,\nabla_s\cdot \bn \,, \quad \bx\in \cal{D}\,,
\end{equation}
concentration $\Gamma$.

The evolution  of the distribution of an insoluble, diffusing surfactant  is governed by a time-dependent convective  equation  \citep{Stone:1990, Wong-Maldarelli}
\begin{equation}
\label{eq:surfactant evolution}
\frac{\partial \Gamma}{\partial t}+\bnabla_{\mathrm{s}}\cdot \left(\bu_{\mathrm{s}}\Gamma\right)+\Gamma \left(\bu\cdot\bn\right)\nabla_{\mathrm{s}}\cdot \bn-D\nabla^2_s\Gamma=0\,\quad{\mbox{at}}\quad r=r_s\,
\end{equation}
where $\bnabla_{\mathrm{s}}$ is the surface gradient operator, $\bnabla_{\mathrm{s}}=\left(\bI-\bn\bn\right)\cdot\bnabla $.

We adopt a linear equation of state for the interfacial tension
\begin{equation}
\label{surfactant equation of state:3}
\gamma(\Gamma)=\gamma_\eq-\left.\frac{\partial \gamma}{\partial \Gamma}\right|_\eq\,\left(\Gamma-\Gamma_\eq\right).
\end{equation}

Henceforth, all variables are nondimensionalized using the radius of the undeformed drops $a$,  the undisturbed field strength $E_0$, a characteristic applied stress $\tau_c=\eps_\out E_0^2$, and the properties of the  suspending fluid. Accordingly, the time scale is $t_c=\eta_\out/\tau_c$ and  the velocity scale is $u_c=a \tau_c/\eta_\out$.  The surfactant concentration is normalized by $\Gamma_\eq$ and the interfacial tension - by $\gamma_\eq$.
 The ratio of the magnitude of the electric stresses and surface tension defines the electric capillary number, the relative strength of the distorting viscous  and restoring Marangoni  stresses is reflected by the Marangoni number and the importance of surfactant diffusion is given by the Peclet number
 \begin{equation}
 \Ca=\frac{\eps_\out E_0^2 a}{\gamma_\eq}\,,\quad\Ma^{-1}=\frac{\eps_\out E_0^2 a}{\Delta\gamma}\,,\quad  \Pe=\frac{\eps_\out E_0^2 a^2}{\eta_\out D}\,.
 \end{equation}
The characteristic magnitude of the surface-tension variations that
result from perturbations of the local surfactant concentration
$\Gamma$ about the equilibrium value $\Gamma_\eq$ is
 \begin{equation*}
\Delta\gamma=-\Gamma_{\rm eq}\left(
\frac{\partial\gamma}{\partial \Gamma}
\right)_ {\Gamma = \Gamma_{\rm eq}}
\end{equation*}
It is convenient to define the elasticity number, which is independent of the externally applied stresses
\begin{equation}
\label{elasticity}
\E=\frac{\gamma_0-\gamma_\eq}{\gamma_\eq}=\Ca\Ma\,.
\end{equation}



{\section{Numerical method}}
We utilize the boundary integral  method to solve for the flow and electric fields. Details of our three-dimensional formulation can be found in \citep{Chiara:2019}.  In brief, the electric field is computed following 
 \citep{Lac-Homsy,Baygents:1998}:
\begin{equation} 
\label{eq:BIE01}
\bE^\infty+\sum_{j=1}^2 \int_{{\cal{D}}_j} \frac{\hat{\bx}}{4\pi r^3} {\left(\bE^\out-\bE^\ins\right)\cdot\bn}dS(\by)= \begin{cases} 
\bE^\ins(\bx)&\mbox{if } \bx $ inside $ \cal{D}, \\ 
\half \left(\bE^\ins(\bx)+\bE^\out(\bx)\right) &\mbox{if } \bx \in\cal{D}, \\ 
\bE^\out(\bx)&\mbox{if } \bx $ outside $ \cal{D}. \\ 
\end{cases}
\end{equation}
where  $\xhat=\bx-\by$ and $r=|\xhat|$. 
The normal and tangential components of the electric field are calculated from the above equation
\begin{equation}
\label{eq:E_n}
\begin{split}
E_n(\bx)&=\frac{2\Rr}{\Rr+1}\bE^\infty \cdot \bn+\frac{\Rr-1}{\Rr+1} \sum_{j=1}^2 \bn(\bx)\cdot\int_{{\cal{D}}_j} \frac{\xhat }{2\pi r^3} E_n(\by)dS(\by)\,,\\
\bE_t(\bx)&=\frac{\bE^\out+\bE^\ins}{2}-\frac{1+\Rr}{2\Rr}E_n \bn\,.
\end{split}
\end{equation}
For the flow field, we have developed the method for fluids of arbitrary viscosity, but for the sake of brevity here we list the equation in  the case of equiviscous drops and suspending fluids. The velocity is given by 
\begin{equation}
 \label{eq:main_eq}
 2\bu(\bx)=-\sum_{j=1}^2  \left( \frac{1}{4\pi}\int_{{\cal{D}}_j} \left(\frac{\bff(\by)}{\Ca}-\bff^E(\by)\right)\cdot \left(\frac{\bI}{r}+\frac{\xhat\xhat}{r^3} \right)dS(\by)\right)\,,
 \end{equation}
where $\bff$ and  $\bff^E$ are the interfacial stresses due to surface tension and electric field
\begin{equation}
\label{eq:interfacial_force}
\bff=\gamma(\bx) \bn\nabla \cdot \bn -\bnabla_{\mathrm{s}} \gamma\,,
 \end{equation}
 \begin{equation}
\label{eq:el_force}
\bff^E=\left(\bE^\out\cdot \bn\right)\bE^\out-\half \left(\bE^\out\cdot\bE^\out\right)\bn-\Sr\left(\left(\bE^\ins\cdot \bn\right)\bE^\ins-\half  \left(\bE^\ins\cdot\bE^\ins\right)\bn\right)\,.
 \end{equation}
For a clean drop, the surface tension coefficient $\gamma(\bx)$ will be constant, and the second term in (\ref{eq:interfacial_force}), the so-called Marangoni force, will vanish.\\
Drop velocity and centroid are computed from the volume averages
\begin{equation}
\bU_j=\frac{1}{V}\int_{V_j}\bu dV=\frac{1}{V}\int_{{\cal{D}}_j}\bn\cdot\left(\bu\bx\right) dS\,,\quad \bx^c_j=\frac{1}{V}\int_{V_j}\bx dV=\frac{1}{2V}\int_{{\cal{D}}_j}\bn\left(\bx\cdot \bx\right) dS\,.
\end{equation}

To solve the system of equations
\refeq{eq:E_n}, \refeq{eq:main_eq}, \refeq{eq:surfactant evolution} we use the Galerkin formulation based on a spherical harmonics representation presented in \cite{Chiara:2019}. In the current study, we update the time scheme to the adaptive fourth order Runge-Kutta introduced in \cite{kennedy2003}. This choice allows to treat the convective term that appear in the surfactant evolution equation \refeq{eq:surfactant evolution} explicitely, and the diffusive term implicitely. To make the implicit part of the solver efficient also for large diffusion coefficients (i.e. Small P\'eclet numbers), a preconditioner designed in \citep{pallson} results to be fundamental to reduce the number of iterations for the convergence. All variables (position vector, velocities, electric field, surfactant concentration etc) are expanded in spherical harmonics which provides an accurate representation even for relatively low expansion order. In this respect, to make sure that all the geometrical quantities of interest (e.g. mean curvature) are computed with high accuracy as well, we use the adaptive upsampling procedure proposed by \cite{rahimian2015}.  A specialized quadrature method for the singular and nearly singular integrals that appear in the formulation and a reparametrization procedure able to ensure a high-quality representation of the drops also under deformation are used to ensure the spectral accuracy of the method \citep{Sorgentone:2018}.


%

Our numerical method and the asymptotic theory for clean drops was presented and validated in \citep{Chiara:2020}. Here we extend the small-deformation theory and the numerical method to include the effect of the insoluble surfactant.\\


\section{Theory: Far-field interactions}
\label{theory}

We first analyze the electrostatic interaction of two widely separated spherical drops. In this case,  the drops can be approximated by  point-dipoles. 
The  disturbance field $\bE_1$ of the drop dipole $\bP_1$ induces a dielectrophoretic (DEP) force on the dipole $\bP_2$ located at $\bx^c_2=d\bR$, given by $\bF(d)=\left(\bP_2\cdot \nabla \bE_1\right)|_{r=d}$ 
The drop velocity  under the action of this force can be estimated from Stokes law, $\bU=\bF /\zeta$, where $\zeta$ is the friction coefficient. For a surfactant-covered drop, $\zeta=6\pi (3\lambda+2+\chi)/(3(\lambda+1)+\chi)$, where $\chi=\Pe \Ma$. Thus, 
\begin{equation}
\label{DEPF}
\bU_2^\dep=2 \frac{\beta_D}{d^4}\left(\frac{\chi+3(1+\visrat)}{\chi+2+3\visrat}\right)\left[\left(1-3\cos^2\Theta\right)\bR-\sin\left(2\Theta\right)\bt\right]\,,\quad \beta_D= \left(\frac{\Rr-1}{\Rr+2}\right)^2
\end{equation}
The velocity reduces to the result for clean drops if $\chi=0$ \cite{Chiara:2020}, and for solid spheres if $\chi\rightarrow\infty$.

\begin{figure}[h]
  \centering \includegraphics[width=0.5\linewidth]{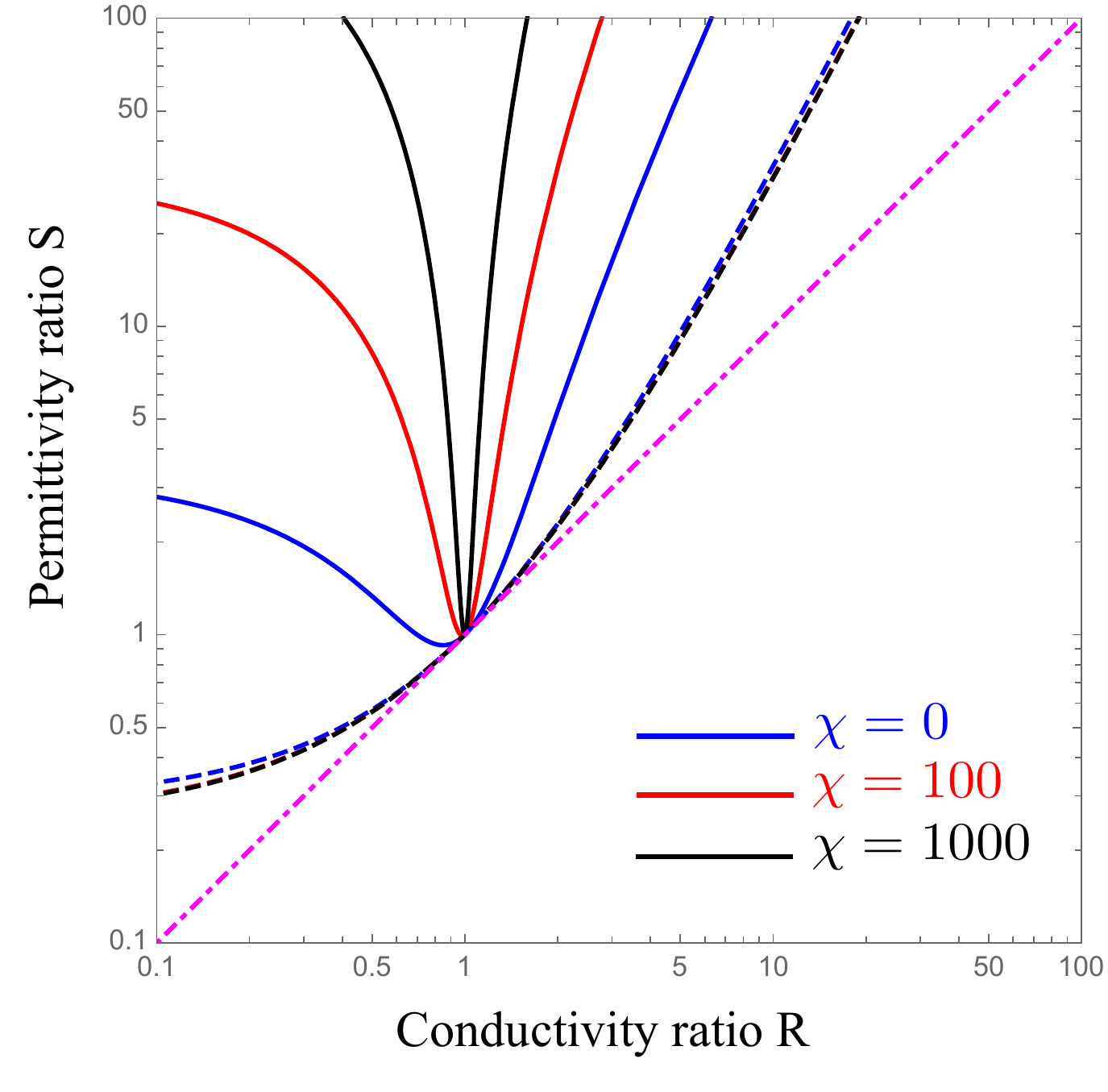}
      \caption{\footnotesize{(a) Phase diagram of drop deformations and alignment with the field for viscosity ratio $\visrat=1$ and different values of the parameter $\chi=\Pe \Ma$.
      The solid lines correspond to $\Phi_s(\visrat, \Rr, \Sr, \chi)=0$ given by \refeq{eq:Phi};  in the parameter space above,
      the line of centers of the two drops rotates away from the applied field direction $\Phi<0$. The dashed lines correspond to the modified Taylor discriminating function \refeq{FT}; in the parameter space above it, drop deformation is oblate and below it - prolate.  Above the dot-dashed magenta line $\Sr=\Rr$, the surface flow is pole-to-equator ($\beta<0$), while below this line the surface flow is equator-to-pole ($\beta>0$) }}
	\label{figT}
\end{figure}

In addition to the dipole-dipole interaction, drops interact hydrodynamically.
Assuming a spherical drop, the electric shear drives a flow, which is 
a combination of a stresslet and a quadrupole \cite{Taylor:1966}
\begin{equation}
\label{ehdU}
 \bu= \frac{\beta}{ r^2}\left(-1+3 \cos^2\theta\right)\rhat-\frac{\beta}{r^4}\left(\left(-1+3 \cos^2\theta\right)\rhat+\sin (2\theta)\that\right)\,.
\end{equation}
The strength of the stresslet is 
\begin{equation}
\beta=\beta_T-\frac{3\Ma}{5(1+\visrat)}g\,,\quad \beta_T=\frac{9}{10}\frac{\Rr-\Sr}{\left(1+\lambda\right)\left(\Rr+2\right)^2}\,,
\end{equation}
where $g$ is  a parameter describing the surfactant redistribution, $\Gamma=1+g\left(-1+3\cos^2\theta\right)$. The surfactant weakens the EHD flow, because the Marangoni stresses due to nonuniform surfactant concentration oppose the shearing electric traction. 
At steady state, the surfactant distribution at leading order is given by the balance of surfactant convection by the electrohydrodynamic (EHD) flow and surfactant diffusion, $\nabla_s\cdot\bu=\Pe^{-1} \nabla^2\Gamma$, which leads to
\begin{equation}
g=\Pe\frac{5(1+\visrat)}{3\left(5(1+\visrat)+\chi\right)}\beta_T\,,
\end{equation}
and thus 
\begin{equation}
\beta=\frac{9\left(\Rr-\Sr\right)}{2\left(\Rr+1\right)^2}\frac{1}{ 5(1+\visrat)+\chi}\,,\quad \chi=\Pe \Ma\,.
\end{equation}
The parameter $\chi$ characterizes the magnitude of the surfactant effect on the EHD flow. In the  limit $\chi=0$ the result reduces to the clean drop solution. In the case of nondiffusing surfactant $\Pe\rightarrow\infty$ ($\chi\rightarrow\infty$), the surfactant completely immobilizes the interface and suppresses the  EHD flow. In this case, the theory predicts that the drops will interact only electrostatically. Moreover, if $\Rr=1$ even the DEP interaction vanishes. Thus a pair of spherical droplets covered with insoluble, nondiffusing surfactant and conductivity ratio $\Rr=1$ will not interact in a uniform electric field. 


The drop translational velocity due to a neighbor drop is found from Faxen's law { \citep{Kim-Karrila:1991, pak_feng_stone_2014}
}\begin{equation}
    {\bU}^\ehd_{2} = \left(1 + \frac{\lambda}{2(3\lambda+2)}\nabla^2\right)\bu|_{\bx=d\bR}\,.
  \end{equation}
Inserting \refeq{ehdU} in the above equation leads to
\begin{equation}
\label{U2ehd}
    {\bU}^\ehd_{2} = \beta\left(\frac{1}{ d^2}-\frac{2}{d^4}\left(\frac{1+3\visrat}{2+3\visrat}\right)\right)\left(-1+3 \cos^2\Theta\right)\bR-\frac{2\beta}{d^4}\left(\frac{1+3\visrat}{2+3\visrat}\right)\sin(2\Theta)\bt+O(d^{-5})\,.
\end{equation}
Combining the electrohydrodynamic and the dielectrophoretic velocities yields
\begin{equation}
\label{U2}
\bU_2=\frac{\beta}{ d^2}\left(-1+3 \cos^2\Theta\right)\bR-\Phi_s\left(\visrat,\Rr, \Sr, \chi\right)\frac{2}{d^4}\left(\left(-1+3 \cos^2\theta\right)\bR+\sin(2\Theta)\bt\right)\,,
\end{equation}
 where
 \begin{equation}
 \Phi_s=\frac{1+3\visrat}{2+3\visrat}\beta+\beta_D\frac{3(1+\visrat)+\chi}{2+3\visrat+\chi}\,.
     \label{eq:Phi}
 \end{equation}
The discriminant ${\Phi_s}$  quantifies the drop pair alignment with the field and the 
interplay of EHD and DEP interactions in drop attraction or repulsion.    Drops with $\Phi_s >0$ move to align their   line-of-centers to with the applied electric field, since $\dot\Theta=\bU_2\cdot \bt\sim -\Phi _s$. If $\Phi_s < 0$ (which occurs only for $\Rr/\Sr<1$ drops),  the line of centers between the drops rotates towards a perpendicular orientation with respect to the applied electric field.  The presence of surfactant reduces the parameter range where misalignment is predicted.
  Figure \ref{figT} summarizes the regimes of alignment and deformation.

The relative radial motion of the two drops at a given separation depends on $\Phi_s$ and $\beta_T$. There is a critical separation $d_c$ corresponding to $\bU_2(d_c)\cdot\bR=0$ at which drop relative radial motion can change sign 
 \begin{equation}
 \label{dc}
 d_c^2=\frac{2(1+3\visrat)}{2+3\visrat}+\frac{\left(\Rr-1\right)^2}{\Rr-\Sr}\left(\frac{4(3(1+\visrat)+\chi)(5(1+\visrat)+\chi)}{9(2+3\visrat+\chi)}\right).
 \end{equation}
 For $\Phi_s > 0$ and $\Rr/\Sr<1$ ($\beta<0)$, $d_c$ does not exist and EHD and DEP interactions are cooperative and act in the same direction (note that system with $\Phi _s<0$ and $\Rr/\Sr>1$ can not exist). For $\Phi_s> 0$ and $\Rr/\Sr>1$ or $\Phi_s<0$ and $\Rr/\Sr<1$, there is competition between EHD and DEP, with the quadrupolar DEP  winning out closer to the drops and the EHD taking over via the stresslet flow in the far-field.  
The critical distance is affected by the presence of surfactant. It increases with $\chi$, since the surfactant weakens the EHD flow and  expands the region of dominance of DEP. In the limit of nondiffusing surfactant, $\chi\rightarrow\infty$, the drop interactions are entirely dominated by DEP.

\section{Results and discussion}
\label{sec:results}
We consider two identical drops with viscosity ratio $\visrat=1$ and focus on the effect of surfactant on drop dynamics under variable $\Rr$, $\Sr$ and initial configuration.

\begin{figure}[h]
  \centering
   \includegraphics[width=\linewidth]{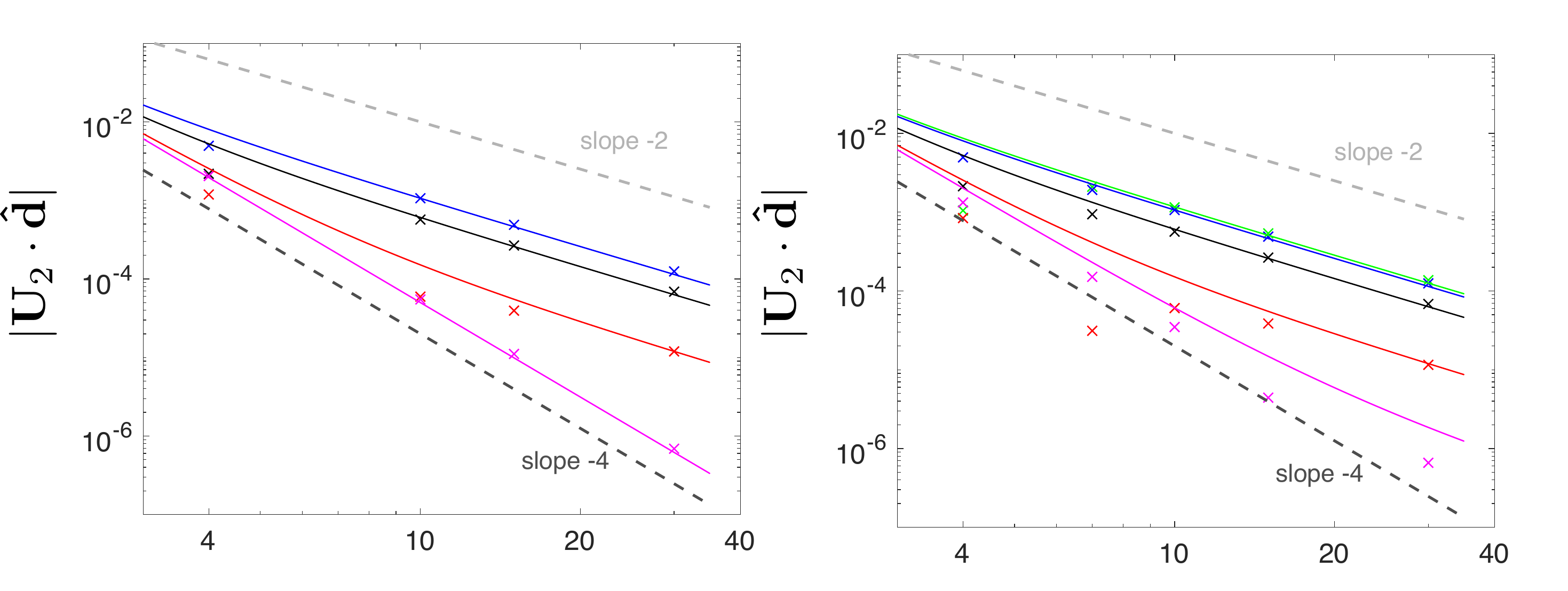}
      \caption{\footnotesize{
      Steady relative velocity of a pair of  leaky dielectric drops aligned with the field ($\Theta=0$).  $\Rr=2$, $\Sr=1$, $\Ca=1$ (left)  $\E=1$, $\Pe=1$ (blue),$\Pe=10$ (black) ,$\Pe=100$ (red) and $\Pe\rightarrow\infty$(magenta). The  symbols are from our fully 3D code and  the solid line is the  theory \refeq{U2}.  In the case of nondiffusing surfactant the interaction is dominated by DEP and the velocity shows $1/d^4$ dependence. (right) $\Pe=1$ and $\E=0,1,10,100,1000$ (green, blue, black, red, magenta). As $\chi=\Ma \Pe$ increases the critical distance beyond which the DEP dominates increases. Note that $\chi=100$ shows change of slope from -4 and -2. $\chi=1000$ slope -4 in the studied range. }}      
	\label{figB}
\end{figure}

First we compare the drop steady velocity obtained from simulations and the asymptotic theory  for a drop pair aligned with the field. 
  Figure \ref{figB} shows that theory and simulations are in excellent agreement, especially at large separations, and the theory is able to capture the steady velocity even for a  relatively  high $\Ca=1$. As the surfactant effect strengthens and $\chi$ increases, either by increase in the surfactant elasticity or decreasing diffusivity, the drops relative velocity switches from  EHD  to DEP dominated at the critical distance \refeq{dc}.  Accordingly the slope dependence on distance changes from $d^{-2}$ to  $d^{-4}$. This is most obvious for the $\chi=100$ case, where $d_c=7.14$. In the limit $\chi\rightarrow\infty$, the drop motion is entirely due to DEP.
\\

\begin{figure}[h]
  \centering  
    \includegraphics[width=\linewidth]{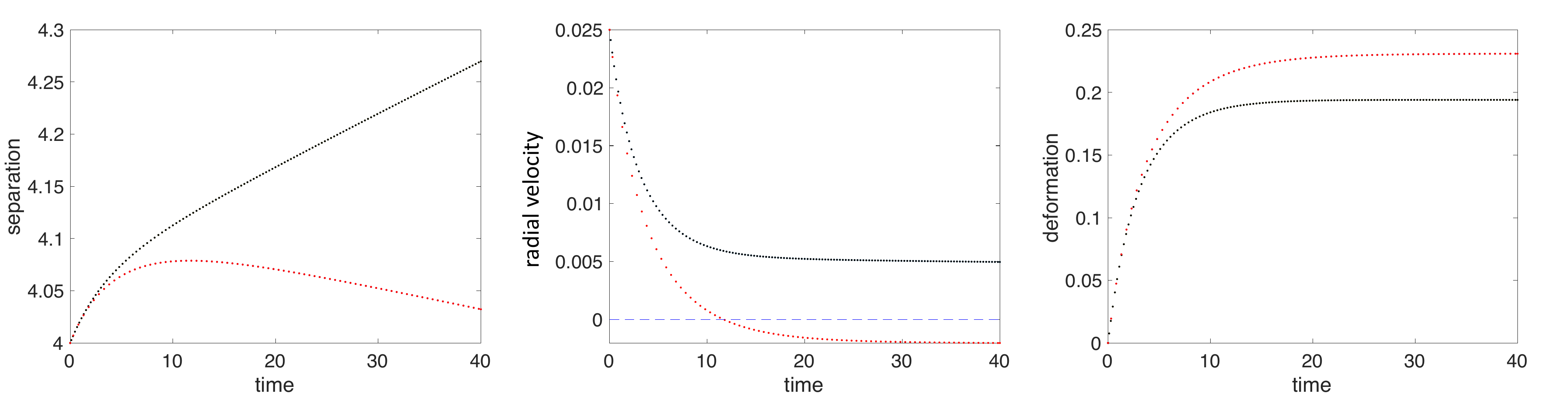} \\
  \caption{\footnotesize{Effect of surfactant on the interaction of two identical drops with $\Rr=2$, $\Sr=1$, $\Ca=1$, $\E=1$ initially aligned with the field $\Theta=0$. Black dots correspond to $\Pe=1$ and red dots correspond to the limit of non-diffusing surfactant $\Pe=10^6$. The surfactant suppresses the electrohydrodynamic repulsion and after initial transient due to shape deformation and surfactant redistribution the interaction can reverse sign. }}
	\label{fig4}
\end{figure}

However, even in this limit where at steady state the interface is immobilized by the surfactant, until the steady  DEP-dominated state is reached, there is EHD affected drop motion due to the transient drop deformation and surfactant redistribution. As a result, the drops can initially repel and then attract once steady drop shape and surfactant distribution are reached. This scenario is illustrated on Figure \ref{fig4}, which shows that the radial relative velocity in the case of a drop covered with non-diffusing surfactant can change sign  from positive (indicating drop repulsion) to negative (attraction). The small-deformation theory which predicts this phenomenon is presented in the Appendix.

\begin{figure}[h]
\centering
\includegraphics[width=\linewidth]{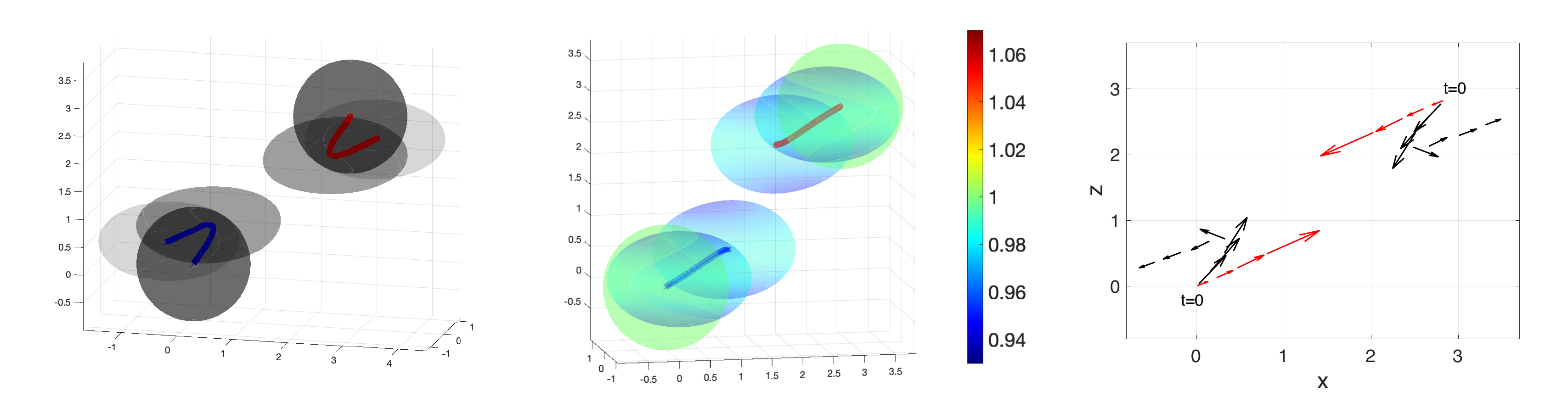}
\caption{$\Rr=0.1, \Sr=5, \Theta=45°$. Initial distance $d=4$.(a) clean drops misaligning (b) non-diffusing surfactant-covered drops with $\E=10$ aligning with the field (c) Center of mass trajectory in the $x-z$ plane. Arrows correspond to the velocity for the clean drops (black) and for the non-diffusing surfactant-covered drops (red). Movies in the additional material.}
\label{fig:fig_al_mis}
\end{figure}

Our previous study of clean drops  \citep{Chiara:2020}  found that  
drops initially misaligned with the field may not experience monotonic attraction or repulsion; instead their three-dimensional trajectories follow three scenarios: motion in the direction of the field  accompanied by either attraction followed by separation or vice versa (repulsion followed by attraction), and attraction followed by  separation in a direction transverse to the field. Next we address the question about the surfactant influence on these intricate dynamics.
The theory presented in Figure \ref{figT} highlighted that the surfactant has two main effects:  first, it increases the range of distances where DEP dominates over EHD, and second, decreases the range of $\Sr$ and $\Rr$ parameters where drops' line-of-centers rotates away from the direction of the applied field. Accordingly,   clean and surfactant-covered drops with same $\Sr$ and $\Rr$, initial configuration and $\Ca$ may display opposite aligning behavior. Figure \ref{fig:fig_al_mis} illustrates such a case.  While the clean drops  attract in the direction of the field and move towards each other, pair up, and then separate in the transverse direction, the surfactant-covered drops only attract and move to align their line-of-centers parallel to the field.

\section{Conclusions}

The effect of surfactant on the three-dimensional interactions of a drop pair in an applied electric field is studied using numerical simulations and a small-deformation theory  based on the  the leaky dielectric model. We present results for the case of a uniform electric field and arbitrary angle between the drops' line-of-centers and  the applied field direction, where the non-axisymmetric geometry necessitates three-dimensional simulations.

The surfactant's main effect is to decrease the electrohydrodynamic flow due to Marangoni stresses compensating the electric shear. As a result, drops' interactions are more strongly affected by DEP: the surfactant-covered drops tend to align with the applied field direction and attract. The surfactant influence is quantified by the parameter $\chi=\Pe\Ma$. The surfactant effect is most pronounced for nondiffusing surfactant ($\Pe\gg 1$) or high elasticity $\Ma\gg1$.  The critical separation at which the DEP overcomes the EHD interaction increases with $\chi$.  The interaction is much weaker compared to the clean drops, because DEP decays with the drops' separation as $1/d^4$  compared to the $1/d^2$ for EHD. The DEP also causes drops to align with the field and the range of $\Rr$ and $\Sr$ where  the drops attract and move in the direction of the field and then separate in the transverse direction is greatly diminished.

\section{Acknowledgments}
PV has been supported in part by NSF award  CBET-1704996.  \\

\appendix
\section{Electrohydrodynamic velocity of a surfactant-covered drop with transient deformation}
Let us consider drop dynamics upon the application of an uniform electric field in the limit of small deformations $\Ca\ll 1$.
At leading order in $\Ca$, the shape and surfactant concentration are  described by $r_s=1+ f(t)\left(-1+3\cos^2\theta\right)$ and  $\Gamma=1+g(t)\left(-1+3 \cos^2\theta\right)$. The deformation parameter is  $D=3/2 f$. 
Combining the small-deformation theories for a surfactant-covered drop in applied flow \cite{Vlahovska:2009a, Vlahovska:Stone} and electric field \cite{Vlahovska_ER:2011,Vlahovska:2019} yields 
\begin{equation}
\dot f=\frac{1}{(3+2\visrat)(16+19\visrat)}\left[15(1+\visrat)t_n^\el+9(2+3\visrat)t_t^\el-\Ca^{-1}\left(4f(10(1+\visrat)+\beta(4+\visrat))-2\beta g(4+\visrat)\right)\right]
\end{equation}
\begin{equation}
\begin{split}
\dot g=&\frac{1}{(3+2\visrat)(16+19\visrat)}\left[9(2+3\visrat)t_n^\el+9(12+13\visrat)t_t^\el-\Ca^{-1}\left(12f\left(2(2+3\visrat)-\beta(8+7\visrat)\right)+6\beta g(8+7\visrat)\right)\right]\\
&+\Pe^{-1} 6 (g-2f)
\end{split}
\end{equation}
where 
\begin{equation}
t^e_n=\frac{1+\Rr^2-2\Sr}{(\Rr+2)^2}\,,\quad t^e_t=\frac{\Rr-\Sr}{\left(\Rr+2\right)^2}
\end{equation}
Steady state deformation depends on the parameter  $\chi=\beta \Pe/\Ca$
\begin{equation}
\label{fS}
f=\frac{3\Ca}{8} F_S\left(\Rr,\Sr, \visrat, \chi\right)\,,
\end{equation}
where  \cite{Ha:1995}
\begin{equation}
\label{FS}
\begin{split}
F_S\left(\Rr,\Sr, \visrat, \chi \right)=&\frac{1}{\left(2+\Rr\right)^2}\left(\Rr^2+1-2\Sr+\left(\Rr-\Sr\right)\frac{3\left(2+3\visrat\right)+2 \chi}{5(\visrat+1)+\chi}\right)\,,
\end{split}
\end{equation}
The limit $\chi=0$ recovers the result for a clean drop $f_{clean}=3F_T/8$, where
$F_T$ is the Taylor discriminating function
\begin{equation}
\label{FT}
\begin{split}
F_T\left(\Rr,\Sr, \visrat\right)=&\frac{1}{\left(2+\Rr\right)^2}\left(\Rr^2+1-2\Sr+3\left(\Rr-\Sr\right)\frac{2+3\visrat}{5(\visrat+1)}\right)\,,
\end{split}
\end{equation}
The limit $\chi\rightarrow\infty$ recovers insoluble surfactant result\cite{Herve:2013}
\begin{equation}
f=\frac{3}{8}\Ca \frac{\left(\Rr+1\right)^2-4 \Sr}{\left(\Rr+2\right)^2}
\end{equation}

The 
 velocity field outside the drop at distance $r$ from the drop center and an angle $\theta$ with the applied field direction  is given by \citep{Vlahovska:Stone}
\begin{equation}
\label{uEHDA}
 \bu=\left(\frac{\alpha+\beta}{ r^2}-\frac{\beta}{r^4}\right)\left(-1+3 \cos^2\theta\right)\rhat-\frac{\beta}{r^4}\sin (2\theta)\that\,,
\end{equation}
where
\begin{equation}
\begin{split}
\alpha=&\frac{15(\visrat+1)}{(3+2\visrat)(16+19\visrat)}\left(F_T\left(\Rr,\Sr, \visrat\right)-\Ca^{-1}\left(\frac{8}{3} f_2(t)+\E\frac{2(4+\visrat)}{15(1+\visrat)}\left(-2f_{2}(t)+g_{2}(t)\right)\right)\right)\\
\beta=&\frac{1)}{(3+2\visrat)(16+19\visrat)}\left(B_T\left(\Rr,\Sr, \visrat\right)-\Ca^{-1}\left(12(2+3\visrat)f_2(t)+\E\left(8+7\visrat\right)\left(-2f_2(t)+g_2(t)\right)\right))\right)\,.
   \end{split}
\end{equation}
where 
\begin{equation}
\begin{split}
B_T\left(\Rr,\Sr, \visrat\right)=&\frac{9 \left(\visrat \left(3 \Rr^2+13 \Rr-19 \Sr+3\right)+2 \left(\Rr^2+6 \Rr-8  \Sr+1\right)\right)}{2 (\Rr+2)^2}\,.
\end{split}
\end{equation}
The shape evolution equation is obtained from the kinematic condition $\dot r_s=u_r(r=1)$. The surfactant evolution is obtained from $\dot \Gamma=-\nabla_s\cdot\bu+\Pe^{-1}\nabla_s^2\Gamma$.

If a second drop is present at location $\bx_2^c = d\bR$, its migration velocity due to the electrohydrodynamic flow of the first drop  can be obtained using Faxen's law \citep{Kim-Karrila:1991}
\begin{equation}
    {\bU}_2^{\ehd} = \left(1 + \frac{\visrat}{2(3\visrat+2)}\nabla^2\right)\bu({r=d}).
\end{equation}
Inserting   \refeq{uEHDA} in the above equation yields
\begin{equation}
\label{U2t}
\begin{split}
U_{2,r}^\ehd&=\left(\frac{\alpha+\beta}{ r^2}-\frac{1}{r^4}\left(\beta+\frac{3\visrat}{2+3\visrat}\left(\alpha+\beta\right)\right)\right)\left(-1+3 \cos^2\theta\right)\\
U_{2,\theta}^\ehd&=-\frac{1}{r^4}\left(\beta+\frac{3\visrat}{2+3\visrat}\left(\alpha+\beta\right)\right)\sin(2\theta)
\end{split}
\end{equation}

At steady state
$\alpha=0$ and $\beta$ reduces to the  result for a spherical drop \refeq{U2ehd}.

\begin{figure}[h]
\centering
\includegraphics[width=\linewidth]{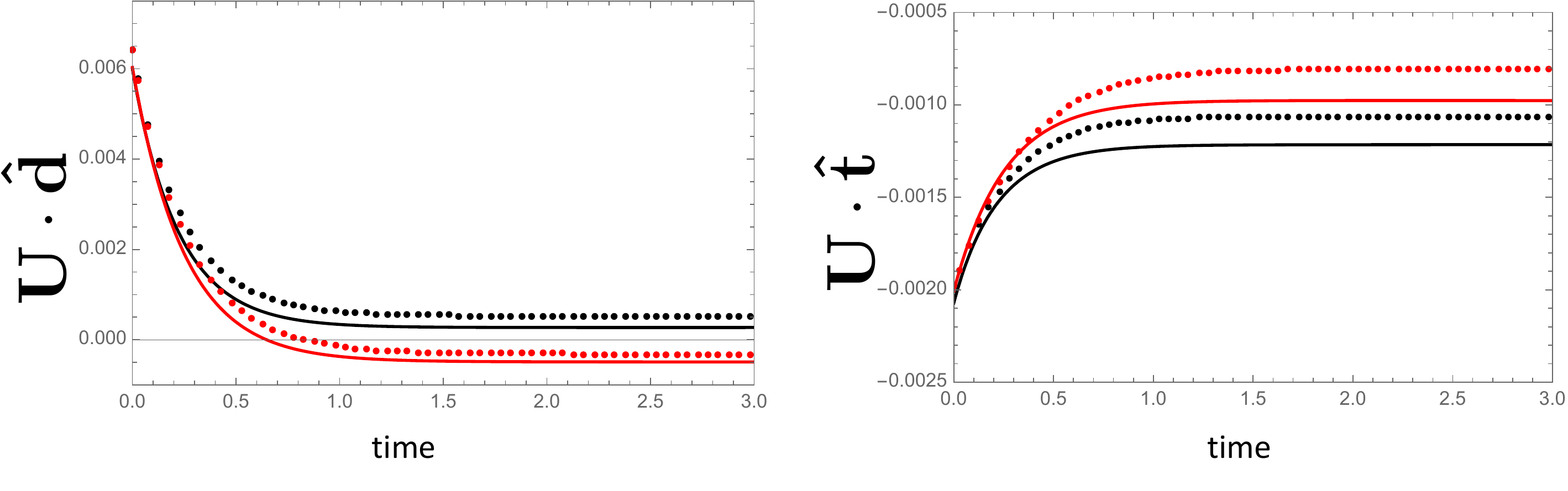}
\caption{\footnotesize{Evolution of the relative radial (left) and tangential (right) velocities for a drop pair with $\Rr=2\,,\,\Sr=1$. Initial angle $\Theta=45°$ and distance $d=4$.  Symbols are numerical simulations and line is the theory. $\chi=1$ (black) and $\chi=10^6$ (red).  Note that the relative radial velocity changes sign for $\chi=10^6$ indicating change from repulsion to attraction. In both cases drops move to aline their line-of-centers with the applied field direction.}}
\label{figA}
\end{figure}

Figure \ref{figA} shows the evolution of the radial and tangential velocity and compares the theory with the numerical simulation.


\section{3D trajectories of surfactant-covered drops in a uniform electric field}

\begin{figure}[h]
    \includegraphics[width=\linewidth]{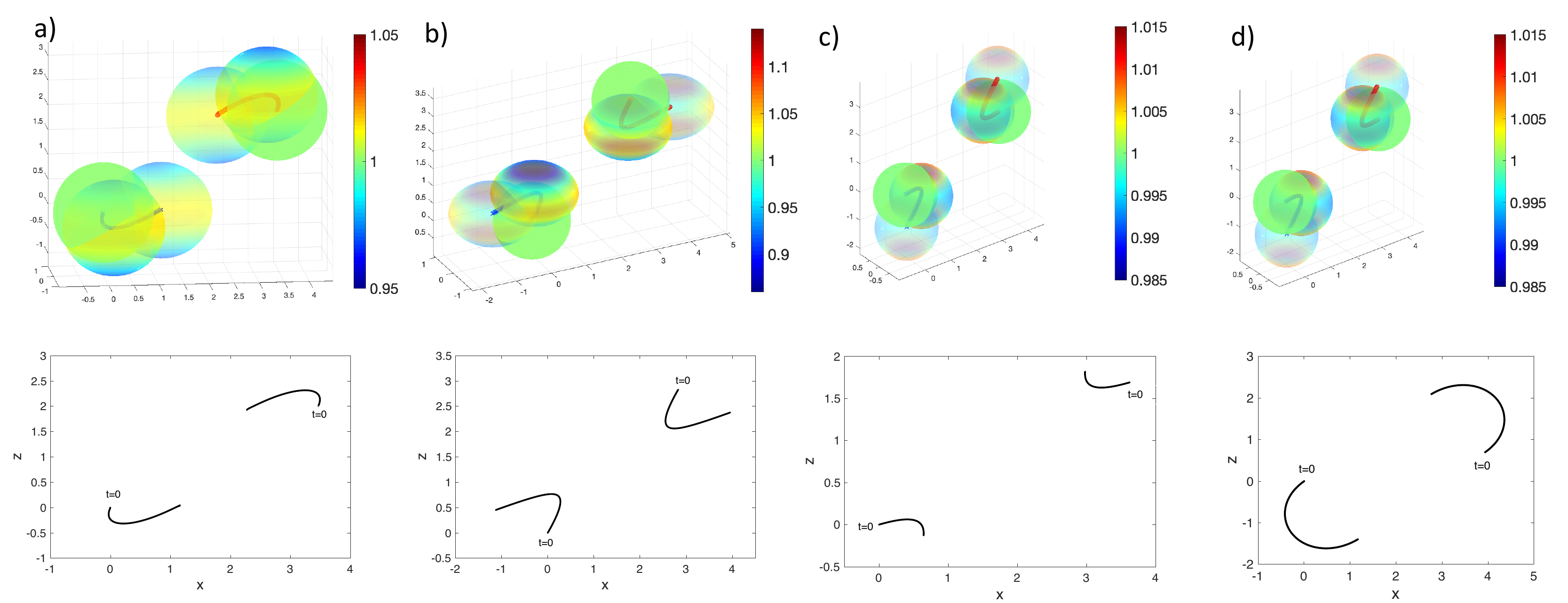} 
  \caption{\footnotesize{Trajectories of two identical surfactant-covered drops with (a) $\Rr=0.1$, $\Sr=1$, (b) $\Rr=1$, $\Sr=10$, (c) $\Rr=1$, $\Sr=0.1$ and (d) $\Rr=100$, $\Sr=1$. Initially the drops are in the xz plane, the separation in all cases is $d = 4$ and the angle with the applied field direction is (a) $\Theta= 60^o$, (b) $\Theta= 45^o$, (c) $\Theta=65^o$, and (d) $\Theta= 80^o$.  $\Ca=0.1$,  $\E=1$ and $\Pe=10^6$. Bottom: trajectories in the xz planes. The color map shows the surfactant concentration}}
	\label{fig9}
\end{figure}

\begin{figure}[h]
  \centering
     \includegraphics[width=\linewidth]{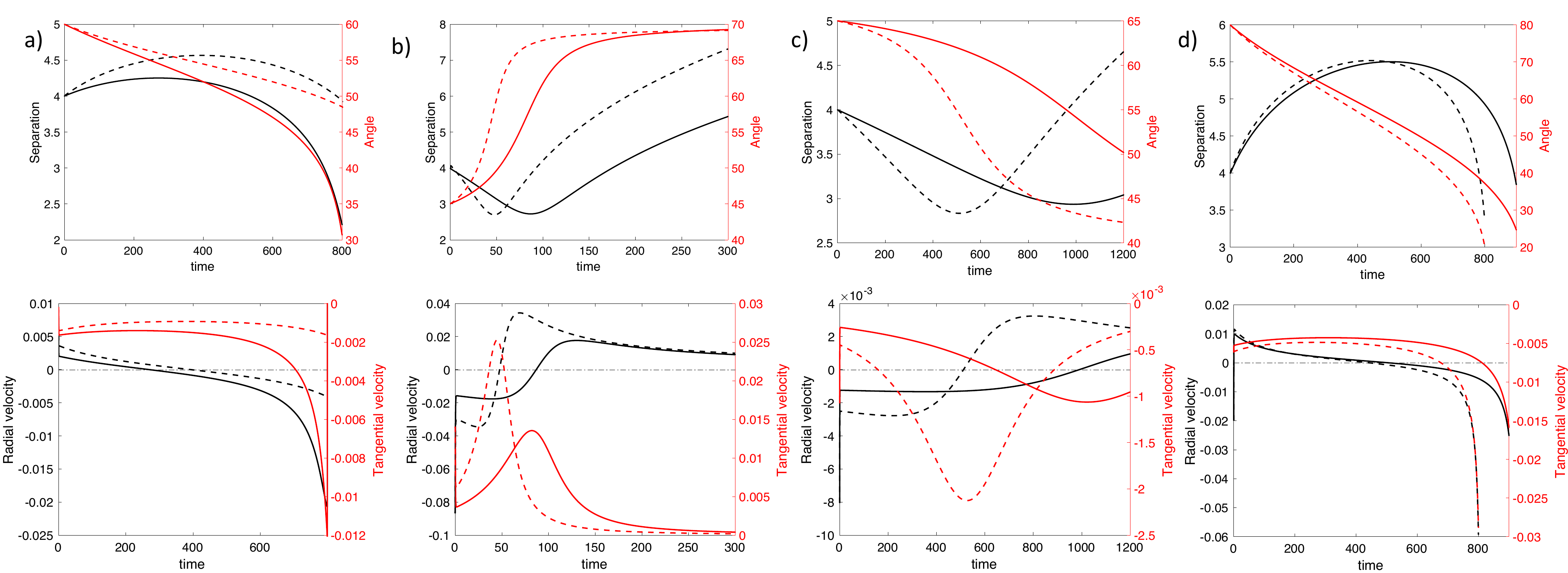}
  \caption{\footnotesize{Dynamics of a pair of identical drops with initial separation $d=4$ and different angles with the applied field. Comparison between clean (dotted line) and surfactant-covered drops (solid line) with {{$\E=1$ and $\Pe=10^6$}}. $\Ca=0.1$.
 (a) $\Rr$=0.1, $\Sr$=1 (repulsion-attraction, alignment with the field), (b)$\Rr$=1, $\Sr$=10  (attraction-repulsion, misalignment with the field), (c) $\Rr$=1, $\Sr$=0.1  (attraction-repulsion, alignment perpendicular to the field), and (d) $\Rr$=100, $\Sr$=1 (repulsion-attraction, alignment with the field).}}
	\label{fig10}
\end{figure}

Next we illustrate  the pair dynamics at different initial configurations. Our previous work showed that clean drops can undergo complex dynamics in an applied uniform electric field if they are initially misaligned with the field: repulsion followed by attraction with centerline rotating towards the applied field direction (a) and (d), attraction followed by repulsion with centerline rotating towards  the applied field direction (c), and  attraction followed by repulsion with centerline rotating away from the applied field direction (b). The drops remain in the plane defined by the initial separation vector and the applied field direction, in this case the $xz$ plane.  The transient pairing dynamics are clearly seen in the trajectories in the $xz$ plane. 
Figures \ref{fig9}-\ref{fig10} show that in these cases the surfactant does not qualitatively change the dynamics, even though the surfactant concentration does become nonuniform. 

\clearpage

\bibliographystyle{unsrtnat}


\end{document}